# $^{75}$As-NMR Studies on LaFeAsO$_{1-x}$F$_x$ ($x$=0.14) under a Pressure of 3GPa

Kenichiro Tatsumi, Naoki Fujiwara*, Hironari Okada[1], Hiroki Takahashi[1], Yoichi Kamihara[2], Masahiro Hirano[2,3], and Hideo Hosono[2,3]

*Graduate School of Human and Environmental Studies, Kyoto University, Yoshida-nihonmatsu-cyo, Sakyo-ku, Kyoto 606-8501, Japan.*

[1]*Department of Physics, College of Humanities and Sciences, Nihon University, Sakurajosui, Setagaya-ku, Tokyo 156-8550, Japan*

[2]*ERATO-SORST, JST, Frontier Research Center, Tokyo Institute of Technology, 4259 Nagatsuda, Midori-ku, Yokohama 226-8503, Japan*

[3]*Frontier Research Center, Tokyo Institute of Technology, 4259 Nagatsuda, Midori-ku, Yokohama 226-8503, Japan*

$^{75}$As-nuclear magnetic resonance (NMR) on an iron (Fe)–based superconductor LaFeAsO$_{1-x}$F$_x$ ($x$=0.14) was performed under a pressure of 3GPa. Enhancement of the superconducting transition temperature ($T_c$) was confirmed from the relaxation rate ($1/T_1$); $T_c$ goes up to 40K by applying pressure up to 3GPa. $1/T_1T$, which is temperature independent just above $T_c$ and gives a measure of the density of states (DOS) at the Fermi energy, enhances by applying pressure. These facts suggest that an increase of the DOS leads to the enhancement of $T_c$. On the other hand, anomalous behavior of $1/T_1T$ observed at high temperatures is suppressed by applying pressure.

Starting from the discovery of LaFeAsO$_{1-x}$F$_x$ system,[1] a series of newly discovered iron (Fe)-based superconductors is currently one of the most exciting topics in strongly correlated electron systems; the superconducting transition temperature ($T_c$) of Sm-, Nd- and Pr-substituted isostructual systems exceeds 50K, which is the first so far other than high-$T_c$ cuprates.[2-5] A variety of compounds including oxygen-free compounds such as Ba$_{1-x}$K$_x$Fe$_2$As$_2$, Li$_{1-x}$FeAs, and α-FeSe [6-8] as well as the oxypnictide superconductors offers various opportunities for theoretical and experimental investigations. Especially, LaFeAsO$_{1-x}$F$_x$ system is an interesting subject of research, because the optimum $T_c$, which is 26K at ambient pressure, goes up to 43K by applying pressure of ~4GPa.[9] Bell-shaped $T_c$ curves have been observed on F doping ($x$) or pressure ($P$) vs. $T$ phase diagram. Controlling these parameters around the optimum conditions can give a clue to rise $T_c$. Especially, pressure is an important parameter because the $T_c$ maximum is realized by applying pressure, whereas enhancement of $T_c$ is not so remarkable for the F doping.[1] To investigate the electric and magnetic properties on a microscopic level, NMR is one of the powerful techniques because the compound includes several kinds of nuclear spins. We performed $^{75}$As ($I$=3/2)-NMR measurements under a pressure of 3GPa and compared to those at ambient pressure.

We used powder samples with the concentration of $x$=0.14, which is located on heavily doped region on the $x$-$T$ phase diagram. The optimum concentration is around $x$=0.11.[1] The $P$ dependence of $T_c$ is shown in Fig. 1. $T_c$ of 19K at ambient pressure, which is a little smaller than that for the $x$=0.11 doped sample, goes up to 43K at the optimum pressure (~5GPa).[10] The value is almost the same with that obtained for the $x$=0.11 doped sample, although a higher pressure is required to achieve the maximum.[11] $T_c$ is also determined using a NMR probe because inductance of the sample coil and thus the resonance condition change remarkably at $T_c$. $T_c$ determined thus is consistent with that obtained from the resistivity measurements.

We operated $^{75}$As-NMR measurements at 45.1MHz using conventional NMR equipment, and pressure was applied using a clamp-type hybrid pressure cell. There appeared three resonance peaks under the field (***H***) corresponding to the transitions, $I^z = \pm 3/2 \Leftrightarrow \pm 1/2$ and $+1/2 \Leftrightarrow -1/2$. The central peak ($+1/2 \Leftrightarrow -1/2$) under pressure is shown in Fig. 2. Broadening of the central peak is explained by the second order perturbation of nuclear quadrupole interaction. The frequency separation between the first satellite peaks $2\nu_Q$ is expressed as $2\nu_Q = e^2QV_{zz}/h$ using the nuclear quadrupole moment ($Q$) and the maximum electric field gradient (EFG), $V_{zz}$ where z axis represents the maximum principle axis. The asymmetry parameter $\eta = (V_{xx} - V_{yy})/V_{zz}$, where $V_{xx}$ and $V_{yy}$ are the EFGs for the other principle axes, is estimated as 0.1 from the spectral line shape at ambient pressure. The value is unchanged under pressure up to 3GPa. The line width $\Delta H$ of the central peak is proportional to $\nu_Q^2$ and $1/H$ [12];

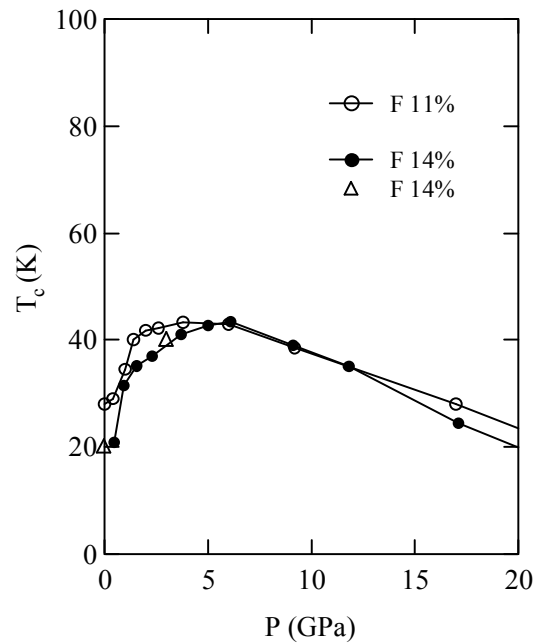

Fig. 1. Pressure ($P$) dependence of the superconducting transition temperature ($T_c$). The concentrations of $x$=0.11 and $x$=0.14 correspond to the optimum and heavily doped region at ambient pressure, respectively. The $T_c$ maximum is almost the same for both samples, although $T_c$ is different at ambient pressure. $T_c$ determined from relaxation rate $1/T_1$ of $^{75}$As is shown as open triangles.

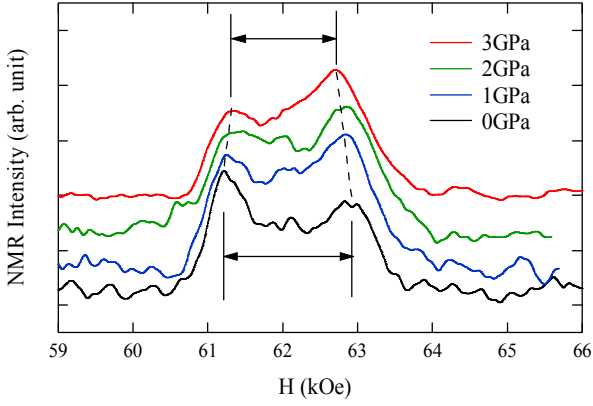

Fig. 2. Central peak corresponding to the transition ($I^z$ =+1/2⇔−1/2) in $^{75}$As ($I$=3/2)-NMR spectra under pressure. The measurements were operated at 30K. The line width shown by arrows is given in Eq. (1) in the text.

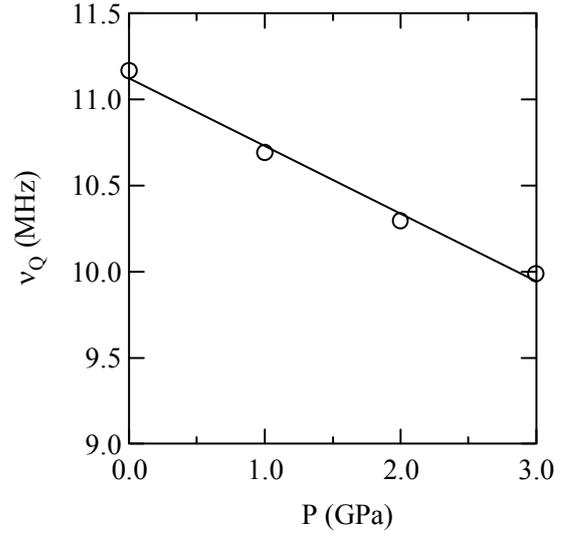

Fig. 3. Pressure dependence of $\nu_Q$ measured from the line width of the central peak

$$\Delta H = \frac{\nu_Q^2}{48\gamma_N^2 H}(25 - 22\eta + \eta^2). \quad (1)$$

In Eq. (1) $\gamma_N$ represents gyromagnetic ratio of $^{75}$As, and is given as 7.292MHz/10kOe. The line width decreases with increasing pressure. As shown in Fig. 3, $\nu_Q$ shows almost $P$-linear dependence and the value changes from 11.13 to 9.95 MHz with increasing pressure from ambient pressure to 3GPa;

$$\nu_Q(MHz) = 11.13 - 0.394P(GPa) \quad (2)$$

In general, nuclear-quadrupole-resonance (NQR) frequency $\nu_{NQR}$ is related to $\nu_Q$ as $\nu_{NQR} = \nu_Q\sqrt{1 + \eta^2/3}$. Because $\eta$ is pressure independent as suggested from the NMR line shape, $P$ dependence of $\nu_{NQR}$ can be attributed to that of $\nu_Q$. The NQR frequency $\nu_{NQR}$ is determined by two factors; one is the EFG originating from the on-site charge density and the other is that from the surrounding Fe ions. The former arises from the hybridization between As-4p and Fe-3d orbitals. The weight of the former and the latter depends on materials. $P$ dependence of the bonding length and angles between the Fe and As ions is not clear for LaFeAsO$_{1-x}$F$_x$ system at present. One may at first suppose that applying pressure causes monotonous shrinkage of the lattice parameters. However, if the bonding length shrinks, $\nu_{NQR}$, namely $\nu_Q$ would enhance without depending on which is the leading factor; the EFG from the surrounding ions increases, and the bonding between the Fe and As ions also becomes strong at the same time, which certainly leads to an increase of the on-site charge density. Therefore, the experimental results are not explained by monotonous shrinkage of the lattice parameters. If the bonding length extends by some reasons, both contributions from the on-site charge and the surrounding Fe ions cause a decrease of $\nu_{NQR}$, namely $\nu_Q$. Such phenomena would be expected only for low pressure region up to the optimum pressure because the bonding length should shrink under extremely high pressure. Although we mentioned only for the bonding length, the bonding angle is actually more important. In any cases, detailed structural information is needed for further analysis.

The relaxation rate $1/T_1$ was measured using a conventional saturation-recovery method. The recovery of the spin echo was fitted by the formula for the central peak of $I$=3/2, $0.1exp(-t/T_1) + 0.9exp(-6t/T_1)$. $1/T_1$ at ambient pressure was already published by several groups.[11), 13-14)] We observed similar $T$ dependence except for behavior at low temperatures. The results of $1/T_1$ are shown in Fig. 4. $T_c$ is determined to be 20K as deviation from $1/T_1T$=constant which is seen just above $T_c$. The relation $1/T_1T$=constant is so called Korringa relation and is proportional to square of the density of states (DOS) at the Fermi energy:

$$\frac{1}{T_1 T} \sim D(\varepsilon_F)^2. \quad (3)$$

Applying pressure up to 3GPa, both $T_c$ and $1/T_1T$ ( = constant) enhance as shown in Fig. 4. The enhancement is clearly seen by $1/T_1T$ vs. $T$ plot ( See Fig. 5). The values of $T_c$ at 3GPa and ambient pressure are shown in Fig. 1. The results are consistent with those observed from the resistivity measurements. Just below $T_c$, $1/T_1$ deviates from the $T$-linear dependence corresponding to the appearance of the superconducting state. A hump of $1/T_1$ corresponding to the coherence peak seems negligibly small. $T^3$ dependence has been observed by several groups.[11, 13, 14] Pairing symmetry has been often discussed from the $T$ dependence in high-$T_c$ cuprates. Theoretical investigations for Fe-based superconductors suggest extended s-wave symmetry[15, 16] or d-wave symmetry.[17] $1/T_1$ follows exponential or power-law function for the former and the latter cases in the clean sample, respectively. In the present measurement $1/T_1$ varies at most one order of magnitude, thus we cannot conclude which is better between exponential and power-law functions.

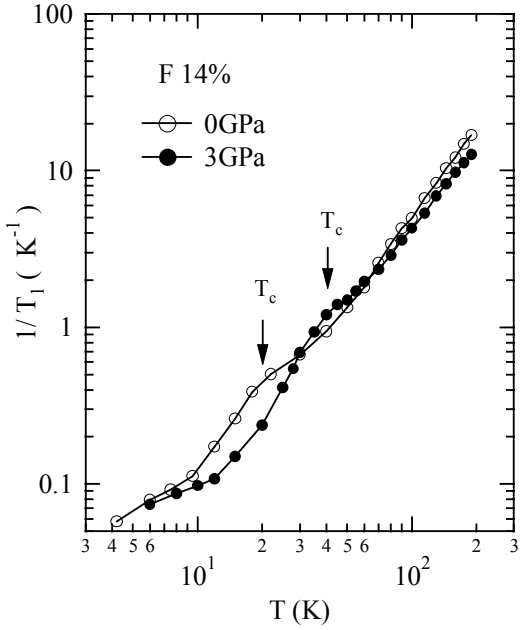

Fig. 4. Relaxation rate $1/T_1$ of $^{75}$As nuclei at ambient pressure (open circles) and 3GPa (closed circle). The curves are guides to eyes.

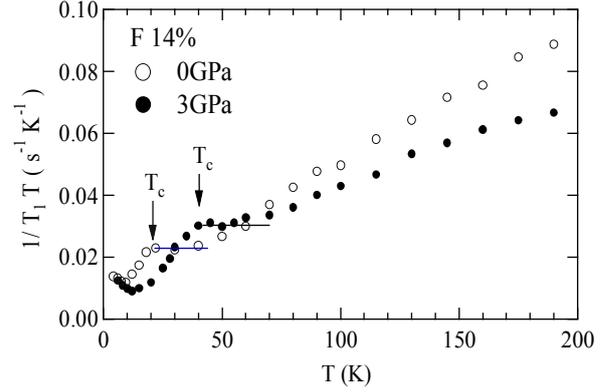

Fig. 5. $1/T_1T$ of $^{75}$As nuclei at ambient pressure (open circles) and 3GPa (closed circle).

At low temperatures below 10K, another relaxation mechanism appears. The $T$ dependence is seen as upturn of $1/T_1T$ at low temperatures in Fig. 5. The mechanism certainly comes from impurity effects. The $T$ dependence has not been reported by other groups.[11, 13, 14] The upturn of $1/T_1T$ may be related to the samples being located on the heavily doped region. The $T$ dependence is not so sensitive to pressure and deviates from $T$-linear dependence which is expected in the unitary limit in the d-wave case. In the case of simple s-wave symmetry, which accompanies no change of sign around the Fermi surface in the gap function, such $T$ dependence is hardly expected. The present results exclude the possibility of simple s-wave symmetry. Recently, $T$-linear dependence or much weaker $T$-dependence has been obtained in the unitary limit for small and large impurity concentration, respectively, in an extended s-wave model.[18] The experimental results may support the possibility of extended s- wave symmetry.

Anomalous behavior of $1/T_1T$ observed at high temperatures is suppressed by applying pressure. If the behavior originated from the energy levels which contribute to the Korringa relation, $1/T_1T$ would increase like that just above $T_c$. Therefore, the origin would come from other energy levels. Some extra energy levels close to the Fermi level may cause the anomalous $T$ dependence. The situation is possible because the present system is a multi-bands system involving five 3d orbitals.

As another possibility, pseudo-gap behavior like high-$T_c$ cuprates may be pointed out.[19] It is worthwhile to compare to high-$T_c$ cuprates. A Curie-like increase of $1/T_1T$ toward $T_c$ has been observed in the case of high-$T_c$ cuprates, and the deviation from the Curie law gives a measure for the pseudo gap. The Curie law comes from the development of $q = (\pi, \pi)$. However, remarkable development of spin fluctuation at some $q$ vectors has not been observed in the present case. In fact $1/T_1T$ shows monotonous decrease with decreasing temperature. $1/T_1T$ at high temperatures seems to show activated-type behavior; $1/T_1T \sim \alpha \exp(-\Delta/T)$. The fitting is effective only for high temperatures and deviates from the data just above $T_c$. The values of $\Delta$ are 98 and 84K at ambient pressure and 3GPa, respectively. The values of $\alpha$ are 0.14 and 0.10 ($s^{-1}$) at ambient pressure and 3GPa, respectively. The result implies that $\Delta$ decreases with increasing pressure and approaches to Fermi-liquid behavior in the whole $T$ range at extremely high pressure. When $1/T_1T$ is fitted by two components, $1/T_1T \sim \alpha_1 + \alpha \exp(-\Delta/T)$, fitting in the whole $T$ range above $T_c$ is possible. The value of $\Delta$, however, increases from 171 to 213K with increasing pressure from ambient pressure to 3GPa. On the other hand, the values of $\alpha$ are 0.16 and 0.12 ($s^{-1}$), and those of $\alpha_1$ are 0.022 and 0.029 ($s^{-1}$) at ambient pressure and 3GPa, respectively. If analogy to high-$T_c$ cuprates holds for this system, $\Delta$ is likely to go down with increasing pressure. Thus, interpretation of $1/T_1T$ by the single exponential function is easier to understand in the pseudo-gap scenario.

Finally we remark a possible scenario for the enhancement of the DOS in relation to the bonding length between the Fe and As ions. The bands which come across the Fermi surface originate from all of the five 3d bands.[20, 21] Finite charge density exists on the As sites through the covalency between the Fe and As ions, although large part of charge density exists on the Fe sites. If the bonding length becomes larger with increasing pressure by some reasons, it causes the narrowing of 3d bands and as a result leads to the enhancement of the DOS at the Fermi energy. The results of $1/T_1T$ as well as $\nu_Q$ can be explained consistently by this scenario.

**Acknowledgment**

We would like to thank K. Kuroki, S. Fujimoto, and M. Sato for fruitful discussions and A. Hisada for experimental support. This work was partially supported by a Grant-in-Aid (KAKENHI 17340107) from the Ministry of Education, Science and Culture, Japan.